\documentclass[twocolumn, secnumarabic,amssymb, aps]{revtex4-1}

\usepackage[columnwise, pagewise]{lineno}

\usepackage[utf8]{inputenc}
\usepackage{CJKutf8}
\usepackage{amsmath,amssymb}
\usepackage{type1cm}
\usepackage{mathrsfs}
\usepackage{mathtools}
\usepackage{url}
\usepackage{graphicx}
\usepackage{color}
\usepackage{booktabs}
\usepackage{braket}
\allowdisplaybreaks[4]

\newcommand\HL[1]{{\color{black}#1}}
\usepackage{bm}
\newcommand{\vect}[1]{\boldsymbol{\mathbf{#1}}}
\def\vec#1{\vect{#1}}

\begin{document}
\title{\HL{Effects of external gravitational field on highly rarefied gases: Analysis based on stochastic soft-sphere collision models}}

\author{Satori Tsuzuki}
\email[Email:~]{tsuzukisatori@g.ecc.u-tokyo.ac.jp\\ \url{https://www.satoritsuzuki.org/} }
\affiliation{Research Center for Advanced Science and Technology, The University of Tokyo, 4-6-1, Komaba, Meguro-ku, Tokyo 153-8904, Japan}
\begin{abstract}
\HL{This study examines the effects of an external gravitational field on highly rarefied gases in the transitional-flow regime near the free-molecular-flow regime. In our theoretical study}, we rederive the classical kinetic theory for an ideal gas in terms of the kinetics of the constituent particles to account for the effect of particle acceleration by an external gravitational field. Subsequently, we derive an extended expression for the virial pressure equation \HL{as a generic description of the dynamics under an external gravitational field}. \HL{We employ the soft-sphere model for the following reasons: In highly rarefied gases, short-range and instantaneous collisional interactions are dominant. Thus, by expanding the asymmetric two-body potential in the virial pressure equation and retaining only the contribution of the short-range interaction, we can obtain a soft-sphere model that represents the interaction in the collision direction as a harmonic oscillation. In the absence of dissipation, the soft-sphere model has been confirmed to reproduce fully elastic collisions}. In our collision simulations, we define two parameters. The first parameter represents the collision probability between each pair of approaching particles, and the second represents the ratio of the magnitude of the external potential energy to the total kinetic energy of the particles. The behavior of the system is analyzed by varying the values of these two parameters. Our analysis shows that if the external potential energy is sufficiently small \HL{(1 $\%$--5 $\%$)} compared with the total kinetic energy, then a pressure difference emerges between the walls. However, the system retains the properties of equilibrium statistical mechanics, as indicated by the Maxwell--Boltzmann (MB) distribution. \HL{In conclusion, highly rarefied gases obey the MB distribution even when placed under weak gravitational fields.}
\end{abstract}
\maketitle

\section{Introduction}
\HL{In classical fluid mechanics, the dynamics of gas flows are characterized by the Knudsen number (Kn), which is a dimensionless parameter defined as the ratio of the mean-free path of molecules to the representative length of the system. In the Kn $<$ 0.001 regime, gas flows can be regarded as continuous flows, as described by the Navier--Stokes (NS) equation. However, in the case of 0.001 $<$ Kn $<$ 0.1, the boundary condition transitions from a nonslip condition to a slip condition, which is referred to as the slip-flow regime~\cite{laurendeau2005statistical, GUO20151}. When Kn $>$ 0.001, the molecular motions become non-negligible and begin to follow the Boltzmann equation~\cite{ESPOSITO20051, ARUMUGAPERUMAL2015955}, which is a kinetic equation that comprises a transport term with respect to a momentum-distribution function on the left-hand side in the differential form, and a collision term in the integral form with respect to momentum and position in the phase space on the right-hand side. Based on the Chapman--Enskog theory~\cite{doi:10.1098/rsta.1916.0006, chapman1990mathematical, Cercignani2002}, the Boltzmann equation converges to the NS equation in the limit where Kn is sufficiently small. Because gas flow depends on multiple independent state variables, the observed flow behavior can change even when in the initial and/or boundary conditions change slightly. Therefore, several opinions exist regarding the threshold value of each regime~\cite{RAPP2017243, BOYD2005669}. Nevertheless, this pertains to the case where the Boltzmann equation can be identified as the governing equation of gas flows for the higher regime of Kn.}

\HL{Directly analyzing the original Boltzmann equation both theoretically and numerically is challenging since it is a nonlinear differential and integral equation. Thus, the Boltzmann equation must be simplified, in particular the collision term involving integral calculations in the phase space for position and momentum, while guaranteeing the conservation of mass, momentum, and energy, and the H theorem~\cite{10.1119/1.10664}. The Bhatnagar--Gross--Krook (BGK)~\cite{PhysRev.94.511, Butta2021} model satisfies these conditions. This model assumes two-body collisions and a local equilibrium distribution function (LEDF), where the collision is represented as a variation of the distribution in the nonequilibrium state from that in the LEDF divided by the relaxation time. This linearizes the right-hand side of the Boltzmann equation, facilitating the numerical simulation of the Boltzmann equation.}
\HL{Meanwhile, as Kn increases, the gas becomes more rarefied. Correspondingly, the heat conductivity of the gas, as a continuum, decreases, and the collective dynamics of the molecules become dominant. In this regime, in addition to calculating the Boltzmann equation using the BGK model, the molecular dynamics (MD) method~\cite{https://doi.org/10.1002/anie.199009921, Nose10061984, HANSSON2002190}, which directly calculates the kinetics of molecules or their samples and obtains macroscopic variables such as pressure and temperature from the ensemble average of microscopic and instantaneous values of physical quantities for each molecule, effectively reproduces the properties of gaseous fluids. The regime in which the Knudsen number is 0.1 $<$ Kn $<$ 10 is known as the transitional-flow regime~\cite{GUO20151}. The Boltzmann equation is the governing equation for this regime. However, the Boltzmann equation assumes the Boltzmann--Grad limit~\cite{Grad1958, Cercignani01011972, Lanford1975, cercignani2012many, 10.1093/imamat/hxr004}, which implies that the ratio of the atomic radius to the mean-free path is sufficiently small. Thus, MD may be more suitable for simulating systems in which the atomic radius is not negligible. Whether it is more accurate to solve the Boltzmann equation or the MD to monitor the dynamics of each molecule depends on the characteristics of the target problem.}

\HL{The collective dynamics of molecules can be essentially described by the pressure relationship~\cite{10.1063/1.2363381, allen2017computer} calculated from the virial theorem. (See Eq.~(\ref{eq:presone}) in Section ~\ref{seq:localpres} for further details), Hereinafter, we refer to this relationship simply as the virial pressure equation (VPE). Briefly, the VPE is derived from Newton's equation of motion for a particle by dividing the total forces exerting on the particle into partial forces from the walls or external forces and those from other particles, and then taking an ensemble average of both sides of Newton's equation of motion. Finally, the VPE has an extended form of the equation of state for an ideal gas. Specifically, its left-hand side represents the pressure, the first term on the right-hand side of the equation represents the contribution to the pressure from the momentum transport due to molecular motion, and the second term on the right-hand side represents the contribution to the pressure from the intermolecular-force interactions. For the latter, the Lennard--Jones (LJ) potential~\cite{doi:10.1098/rspa.1924.0081, doi:10.1098/rspa.1925.0147, PhysRevA.2.221} is widely acknowledged, which represents the interaction potential energy between molecules derived from the chemical and physical properties of electron clouds and molecule dipoles. Additionally, in extremely rarefied gases with large Kn, such as Kn $>$ 10, interparticle collisions are negligible, and particle-wall collisions (and hence the wall boundary conditions) dominate the macroscopic properties of the system~\cite{laurendeau2005statistical}. In this regime, the molecules can be regarded as almost-free particles, and because intermolecular interactions are negligible, they can be assumed to propagate in a straight line until they collide with the walls. Accordingly, the region where Kn $>$ 10 is known as the free-molecular-flow regime.}

\HL{This study aims to investigate the effects of weak external gravitational fields on highly rarefied gases in the regime where Kn is sufficiently high, e.g., Kn $\approx$ 10, which corresponds to the regime from the upper bound of the transitional-flow regime to the lower bound of free-molecular-flow regime. Here, we target the ideal case of highly rarefied gases, and the ``external gravitational field'' represents a general gravitational field, which may refer not only to the Earth's gravitational field but also to those of other planets~\cite{fleagle1981introduction, merriam1992atmospheric} or to external electric fields imposed on charged particles that interact only when they collide owing to their high-energy state. Additionally, ``weak'' in this context refers to cases where the external potential energy is sufficiently small (1 $\%$--5 $\%$) compared with the total kinetic energy of the particles. We employed a numerical approach, as it allows us to investigate a wide range of parameters, unlike actual experiments. Our numerical approach for reproducing the gas flow in this regime is summarized below. Recall that related studies have primarily discretized and simulated the Boltzmann equation up to the intermediate level of rarefied gases ~\cite{Ghiroldi_Gibelli_2015, YUAN201625, 10.1115/1.4031000}. However, in an extremely rarefied gas with a high Kn, the distribution function loses its smoothness and approaches a discontinuous function. In fact, the Lattice Boltzmann method~\cite{Kruger2017, PhysRevE.56.6811, 501535e337e94fe89cef97bacb43f167}, which solves the Boltzmann equation using the BGK approximation model, can effectively reproduce coarse-grained mesoscale gas flows from the perspective of classical statistical mechanics; however, it does not directly capture the microscopic behavior of molecules. Furthermore, it does not easily satisfy the boundary conditions for complex geometries with adequate accuracy. By contrast, the Lagrangian approach, which monitors the motion of molecules whose physical properties are defined by themselves, is promising.}

\HL{Next, we consider MD approaches. Calculating the kinetic dynamics of molecules using a first-principles approach is unrealistic. However, one can solve the N-body interaction system for a group of molecules sampled from all molecules as an alternative or focus on a small volume of the region of interest containing a finite number of molecules. In fact, we can represent the system using VPE in both cases; nevertheless, we accept the former interpretation regarding physical significance for the following reasons. In rarefied gases in the transition regime, which is close to the free-molecular-flow regime, interactions between particles occur only when they collide with each other. These interactions are instantaneous. This implies that the dissipation during the interaction is negligible, thus allowing us to assume fully elastic collisions with a coefficient of restitution $e = 1$; a similar assumption is presented in Ref.~\cite{bird1994molecular}. Moreover, let us assume that the internal degrees of freedom of the molecular particles, i.e., rotational effects, are negligible. Consequently, the particles would receive repulsive forces only in the direction normal to the collision. In this limited case, the distinction in the particle size resulting from the discrepancy between the two interpretations significantly affects the frequency of collisions or, in other words, the collision cross-section (CCS) of the particles. However, at the zeroth-order approximation level, we can assume no significant discrepancy between the two interpretations with respect to other characteristics, such as the scattering-angle distribution, provided that sufficient statistical data and randomness are ensured. Nonetheless, the potential necessity of incorporating quantum mechanical effects when targeting a small domain of interest on an actual scale should be further considered. Consequently, in the context of classical mechanics, gas flows should be modeled based on the larger of the two scales, as in the former interpretation. As mentioned previously, the CCS of each particle differs between the first and second interpretations; in other words, the collision frequencies between the two particles are different. In this regard, we provide the CCS as an input parameter to the simulation system. Specifically, to determine the collision frequency, we define the collision probability between particles as a constant input parameter to the system and then analyze its characteristic behavior by changing its value. }

\HL{For collisions, the effect of the symmetric two-body potential between particles $i$ and $j$ is negligible if particle $j$ is outside the neighborhood of particle $i$. This is because they behave as free molecular particles in highly rarefied gases when they do not collide with other particles. Hence, unlike the LJ or soft-sphere potential~\cite{Dufty2994, PhysRevE.49.1251, 10.1063/1.3266845} in the MD framework, two types of collision models may be suitable for reproducing highly rarefied gases: (1) hard-sphere models~\cite{10.1063/1.1730376, ALLEN1989301, 10.1063/1.858656, 1997529, Isobe01112016}, which assume an infinite potential inside a rigid sphere and otherwise zero (they are well-acknowledged collision models in direct Monte Carlo simulations); and (2) soft-sphere models~\cite{Wassgren2006, BUIST2016363, Zhou2024}, which allow particles to penetrate each other and describes the motion between two particles in the collision direction as a harmonic oscillator. As described in Section~\ref{seq:localpres}, a Taylor-series expansion of the symmetric two-body potential function, which results in only the terms contributing to the collisional interaction, is known to yield a soft-sphere model. Accordingly, the soft-sphere model conforms better to MD theory than the hard-sphere model. Therefore, we adopted a soft-sphere model to simulate the effect of a weak external gravity field on a highly rarefied gas.}

\HL{The remainder of this paper is organized as follows: In the Methods section, we first review the detailed expressions of the classical kinetic theory of gases (CKTG)~\cite{grad1958principles, Boltzmann1964, BALMER2011727, BIRD1993442} for ideal gases. As mentioned previously, the VPE can be segregated into momentum transport and many-body interaction terms. The former corresponds to the total kinetic energy of free particles in an ideal gas. We rederive the first term by assuming that external gravitational fields accelerate these particles by modifying the CKTG for an ideal gas; thus, as a preliminary step, we briefly review the original CKTG for an ideal gas. Additionally, we review the VPE. Subsequently, we obtain a modified version of the VPE that considers the effect of the external gravitational field on the microscopic kinetics of the particles. The modified VPE predicts that if the external potential energy of each particle is sufficiently small (not negligible) compared with its kinetic energy, then the pressure increases in proportion to the distance from the reference point along the direction of gravity. By contrast, for the second term in the VPE, i.e., the many-body interaction term, the interaction force between two particles can be expressed as a soft-sphere model if we retain only the contributions of the short-range (collision) interactions, which occur as the particles approach each other in the Taylor expansion of the symmetric two-body potential. More importantly, the soft-sphere model in this context is a spring-force model that considers the harmonic potential between two particles. The discrete-element method (DEM)~\cite{Cundall1979, doi:10.1080/19648189.2008.9693050, DANBY2013211} is a representative contact-force model that assumes a spring force with dissipation in both the normal and tangential directions against the contact direction. In fact, the soft-sphere model used in this study corresponds to a specific case of the DEM that considers the interactions between particles only in the normal direction without dissipation. Therefore, the spring coefficient and appropriate time can be selected by referring to the DEM settings. We performed several multiparticle collision simulations using the spring-force model, as detailed in the Numerical Analysis section. We discuss the simulation results for two-dimensional (2D) benchmark cases with different values of two characteristic parameters: (i) the ratio of the external potential energy to the kinetic energy in each particle and (ii) the probability of particles colliding with each other.}

\HL{In the Discussion section, we review and discuss the obtained findings: (a) If the potential energy specified by an external gravitational field is sufficiently small (but not negligible, i.e., 1 $\%$--5 $\%$ of the total kinetic energy), then the gas pressure increases with the distance from the gravitational reference point, as predicted by our model. However, the pressure difference exerted by such a weak gravitational field does not perturb the equilibrium statistical properties of the system, which shows the Maxwell--Boltzmann (MB) distribution in the direction of gravity. (b) In the absence of interactions between particles, the system exhibits deterministic behavior. However, as the frequency of interactions increases, the system changes from deterministic to statistical behavior and begins to reflect equilibrium statistical mechanics, which is presented by the MB distribution. (c) During the transition from deterministic to statistical behavior, the system exhibits unstable behavior. (d) Results of (b) and (c) independent of the external gravitational field. Notably, as shown in (a), the pressure difference exerted by a sufficiently small external potential field does not diminish the equilibrium statistical nature of the system. Instead, during the transition from a deterministic to a statistical state, the addition of an external gravitational field stabilized the system and facilitated the MB distribution. This implies that highly rarefied gases obey the MB distribution even when placed in weak gravitational fields. In this section, the physical properties of the simulation results are discussed, and the results are summarized.}

\begin{figure*}[t]
\vspace{-32.5cm}
\hspace{+1.8cm}
\includegraphics[width=4.15\textwidth, clip, bb= 0 0 4405 2480]{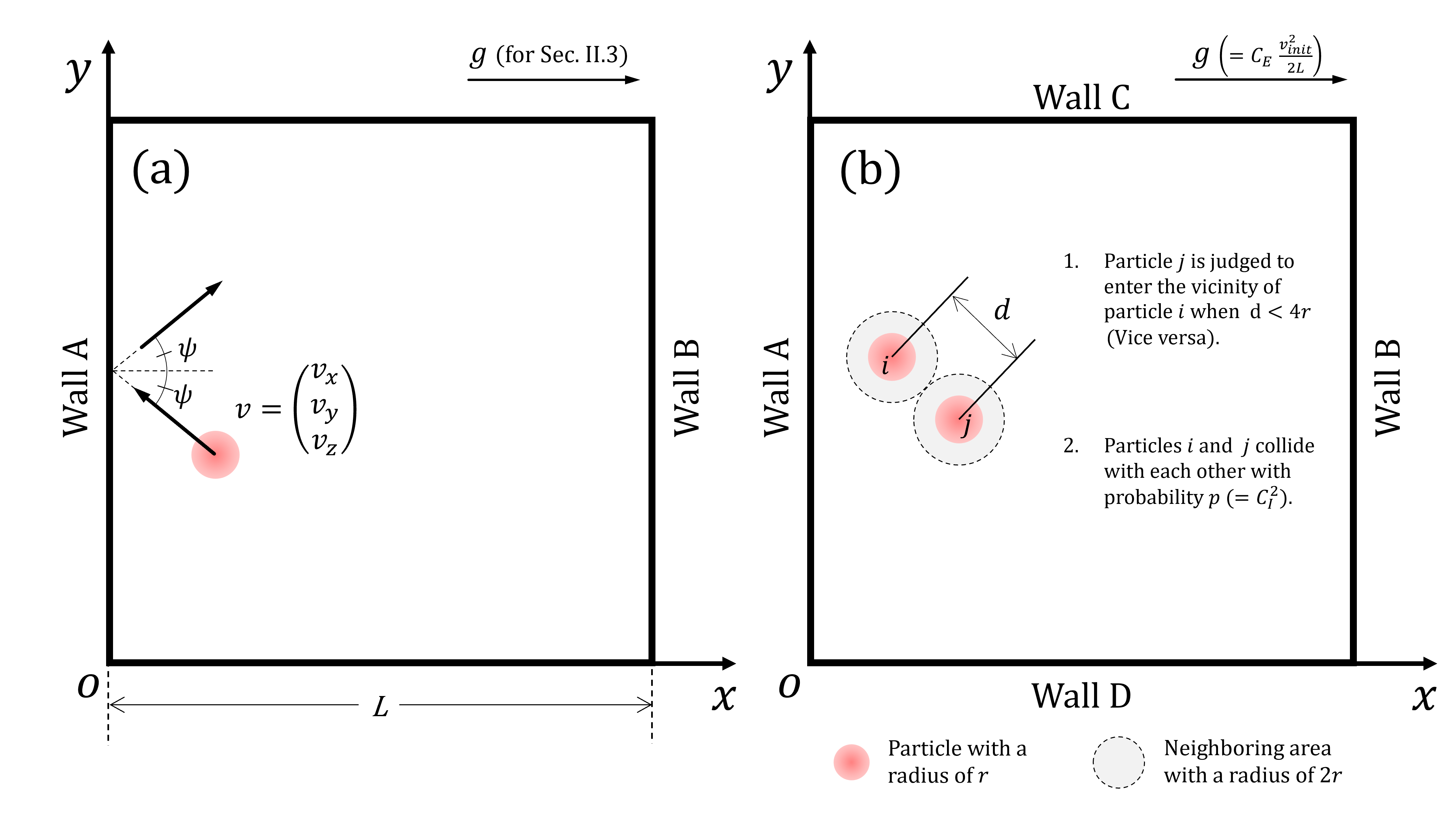}
\caption{\HL{Schematic illustrations of problem setup in (a) Section~\ref{sec:methods}, which presents the classical kinetic theory for an ideal gas, and that in (b) Section~\ref{sec:NumAnalysis}, which presents the interaction model with the collision probability, as used in our simulations}}
\label{fig:Figure1}
\end{figure*}

\section{Methods} \label{sec:methods}
\subsection{\HL{Brief review} of classic kinetic theory of gases \HL{for ideal gas}}\label{seq:CKTGexpl}
The CKTG describes the thermodynamic properties of a gas based on the microscopic motions of its constituent molecules~\cite{grad1958principles, Boltzmann1964, BALMER2011727, BIRD1993442}. Initiated by Bernoulli, the CKTG was subsequently developed by eminent physicists, including Clausius and Maxwell. Consequently, physicists introduced a many-particle microscopic mechanical perspective into thermodynamics based only on fundamental knowledge of physics. For an ideal gas, the characteristics of CKTG can be summarized as follows: (1) The molecules are in random motion, and the square of the root mean square (RMS) velocities are identical in all directions; (2) the molecular volume is sufficiently small to be negligible; (3) the interactions between molecules are negligible; (4) the kinetic energy and momentum are fully conserved in collisions between molecules; and (5) the kinetic energy and momentum are conserved in collisions between the molecules and container walls. \HL{Unless otherwise indicated, the aforementioned conditions (1)--(4) will be collectively referred to as the ``condition (X) for an ideal gas'' hereinafter. Conditions (4) and (5) imply that the coefficient of restitution $e$ between the particles and between the particles and a wall is equal to 1. Here, the $e$ between particles $i$ and $j$ is defined as $e = |\acute{v}_{i}-\acute{v}_{j}|/|v_{i}-v_{j}|$, where $\acute{v}_{i}$ and $\acute{v}_{j}$ represent the velocities of particles $i$ and $j$ after collision, and $v_{i}$ and $v_{j}$ represent their velocities before collision. When particle $i$ collides with a fixed wall, the $e$ is determined by setting the velocities of the remaining interacting particles to zero: $v_{j} = \acute{v}_{j} = 0$. In multidimensional cases, these relationships are applied in each direction.} 

Consider a cube of length $L$ in each direction, wherein an ideal gas comprising $N$ constituent particles is enclosed. No gravitational field is applied to the system. The velocities of a particle and its components are denoted as $v$, $v_{x}$, $v_{y}$, and $v_{z}$. Consider a scenario in which a particle traverses between two walls, A and B, which are parallel to the $x$-direction. Wall A is located to the left of wall B. \HL{Figure~\ref{fig:Figure1}(a) shows an overview of the problem setting discussed in Section ~\ref{sec:methods}, where only one particle is depicted among the N particles for ease of explanation.} For our discussion, the direction of the vector is positive to the right (from wall A to wall B). Vector quantities such as forces and velocities are not bolded in this section because we only consider one-dimensional directions between walls A and B.
\HL{Note that parameters $g$ and $\psi$ in Fig.~\ref{fig:Figure1}(a) are used only in Sections~\ref{sec:modkineg} and~\ref{seq:discuss}, respectively, and are not referred to in this section.}
We denote the impulse exerted on wall B during a collision as $I$. Because the impulse provided to a wall during a collision is equivalent to the momentum variation before and after the collision, we can express $I$ as $2 m v_{x}$, where $m$ is the particle mass.
Because the time interval between the first and second collisions on wall B is expressed as $\frac{2L}{v_{x}}$, the total impulse exerted on wall B by the particle during a discrete time $\Delta t$, $I_{tot}$ is expressed as $I_{tot} = \frac{mv_{x}^2 \Delta t}{L}$. 

The force exerted by a particle on wall B in units of time is denoted by $f$. Based on the definition of an impulse, we have $I_{tot} = f \Delta t$. A comparison of these expressions for $I_{tot}$ yields $f = \frac{m v_{x}^{2}}{L}$. As in condition (3), for an ideal gas, the effects of interactions among particles are assumed to be negligible and each particle is free. Accordingly, the total force received by wall B from $N$ particles, $F$, is expressed as $F = \frac{N m \overline{v_{x}^{2}}}{L}$, where $\overline{X}$ represents the ensemble average of $X$. Owing to the isotropy $\overline{v_{x}^{2}} = \overline{v_{y}^{2}} = \overline{v_{z}^{2}}$, we obtain the relationship $F = \frac{N m \overline{v^{2}}}{3L}$. The pressure $P$ is defined as the force $F$ per unit area, i.e., $P \coloneqq \frac{F}{L^{2}}$; and $P$ on wall B can be expressed as $P = \frac{N m \overline{v^{2}}}{3V}$, where $V$ is a volume of size $L^3$. However, the equation of state (EOS) for an ideal gas is $P V = N k_{B} T$, where $k_{B}$ is Boltzmann's constant, $N$ the number of particles, and $T$ the temperature. Accordingly, the internal energy of the system, $U$, can be described in two different forms: $U = \frac{N m \overline{v^{2}}}{2} = \frac{3}{2} N k_{B}T$. Consequently, the temperature $T$, which is an intensive variable of the system, can be quantitatively described as the sum of the kinetic energies of the individual constituent particles.

\subsection{Expressions of local pressure in $N$-particle interacting systems} \label{seq:localpres}
Next, we explain the relationship between the CKTG \HL{for an ideal gas} and MD to confirm that an MD gas is expressed as an extension of an ideal gas by considering the effects of interactions between particles.
In MD, the thermodynamic pressure $P$ exerted by $N$-interacting particles in a closed domain can be expressed as an ensemble average of the local pressure $\mathcal{P}$ in time and space as $P \coloneqq \braket{\mathcal{P}}$~\cite{gray1984theory, 10.1063/1.2363381, allen2017computer}. The expression for $\braket{\mathcal{P}}$ is obtained as follows: The force exerting on the $i$-th particle can be decomposed into the subforce obtained as the sum of the interaction forces from the other particles, $\vec{F}_{int}^{(i)}$, and that from the walls, $\vec{F}_{wall}^{(i)}$. Accordingly, the equation of motion is expressed as $m \ddot{\vec{r}}_{i} = \vec{F}_{int}^{(i)} + \vec{F}_{wall}^{(i)}$. Here, $\ddot{X}$ is the second derivative of $X$ and $\vec{r}_{i}$ is the position vector of the $i$th particle. By taking the inner product of vector $\vec{r}_{i}$ for both sides, while considering the long-time average for both sides of the equation, followed by summing for $N$ particles, we obtain the following equation after applying the relationship $\frac{N}{2}m \overline{v^{2}} = \frac{3}{2}N k_{B} T$ and assuming that the time and spatial averages are commutative~\cite{10.1063/1.2363381, allen2017computer}: 
\begin{eqnarray}
\braket{\mathcal{P}} &=& \frac{1}{V} \biggl[ N k_{B} T + \frac{1}{3}\sum _{i} \vec{r}_{i} \cdot \vec{F}_{i} \biggr]. \label{eq:presone}
\end{eqnarray}
Here, the relationship $\braket{\sum _{i} \vec{r}_{i}\cdot \vec{F}_{wall}^{(i)}} = -3PV$ is used, as described in~\cite{allen2017computer}. This is because the force of the $i$th particle exerting on a wall is expressed as an anti-force of $\vec{F}_{wall}^{(i)}$, and the pressure $P$ is defined as the force per unit area. The ensemble average of $\sum_{i} \vec{r}_{i}\cdot \vec{F}_{wall}^{(i)}$ over time and space, $\braket{\sum_{i} \vec{r}_{i}\cdot \vec{F}_{wall}^{(i)}}$, can be replaced with the surface integral of $P$ over the wall as $\braket{\sum_{i} \vec{r}_{i}\cdot \vec{F}_{wall}^{(i)}} = -\int P \vec{n}\cdot \vec{r} dS$. By focusing on pairwise interactions, Eq. ~(\ref{eq:presone}) can be further rewritten as~\cite{10.1063/1.437577, 10.1063/1.2363381, allen2017computer}
\begin{eqnarray}
\braket{\mathcal{P}} &=& \frac{1}{V} \biggl[ N k_{B} T + \frac{1}{3}\sum _{i=1} \sum _{j<i} \vec{r}_{ij} \cdot \vec{F}_{ij} \biggr]. \label{eq:prestwobodyf}
\end{eqnarray}
Here, $\vec{r}_{ij}$ is the relative vector that satisfies $\vec{r}_{ij} \coloneqq \vec{r}_{i} - \vec{r}_{j}$, where $\vec{r}_{i}$ and $\vec{r}_{j}$ are the position vectors of the $i$th and $j$th particles, respectively. Furthermore, $\vec{F}_{ij}$ represents the interparticle forces between the $i$th and $j$th particles.

Consider the case where $\vec{F}_{ij}$ is provided as the partial derivative of the symmetric two-body potential $\phi$ with respect to $\vec{r}_{i}$ as $\vec{F}_{ij} = - \frac{\partial \phi(r_{ij})}{\partial \vec{r}_{i}}$, with $r_{ij} = |\vec{r}_{ij}|$ and $\frac{\partial \phi}{\partial \vec{r}} \coloneqq (\frac{\partial \phi}{\partial x}, \frac{\partial \phi}{\partial y}, \frac{\partial \phi}{\partial z})$. Based on simple calculations, Eq.~(\ref{eq:prestwobodyf}) can be written as ~\cite{10.1063/1.437577, allen2017computer}
\begin{eqnarray}
\braket{\mathcal{P}} &=& \frac{1}{V} \biggl[ N k_{B} T - \frac{1}{3}\sum _{i=1} \sum _{j<i} r_{ij} \frac{d \phi(r_{ij})}{d r_{ij}} \biggr], \label{eq:prestwobodypot}
\end{eqnarray}
where we utilize the following mathematical relationships: $\frac{\partial \phi(r_{ij})}{\partial \vec{r}_{i}} = \frac{\partial r_{ij}}{\partial \vec{r}_{i}}\frac{d \phi(r_{ij})}{d {r}_{ij}}$ and $\vec{r}_{ij} \cdot \frac{\partial r_{ij}}{\partial \vec{r}_{i}} = \vec{r}_{ij} \cdot \nabla r_{ij} = r_{ij}$. This is a generic expression for the ensemble average of the local pressure for $N$-particle interacting systems under the symmetric two-body potential required in this study.

Consider the case where $\phi$ is minimized and becomes zero \HL{when} $r_{ij}$ \HL{corresponds to an equilibrium point} $d$ and the derivative of $\phi$; that is, the interaction force $\vec{F}_{ij}$ becomes zero. At this time, $\phi$ satisfies $\phi (d) = \phi^{\prime} (d) = 0$. Consequently, we can expand $\phi$ around $d$ in the Taylor series using $\theta = r_{ij}-d$~\cite{Greiner2001}:
\begin{eqnarray}
\phi(r_{ij}) &=& \phi(d) + \phi^{\prime}(d)\theta 
                 + \frac{\phi^{\prime\prime}(d)}{2 !}\theta^{2} + \mathcal{O}(\theta^{3}), \nonumber \\
		&\therefore& \phi(r_{ij}) \approx \frac{k}{2}\theta^{2}, \label{eq:tayloroscires} 
\end{eqnarray}
where $k = \phi^{\prime\prime}(d)$. Accordingly, $\phi$ denotes the harmonic-oscillation potential. Furthermore, $k$ can be referred to as the spring coefficient. From a mathematical perspective, the Taylor expansion above assumes that the symmetric two-body potential $\phi$ is infinitely differentiable within the closed interval $[d,~r_{ij}]$. Additionally, the effect of the higher-order terms of order $n \ge 3$ is disregarded in the final portion of Eq.~(\ref{eq:tayloroscires}). The Taylor expansion in Eq. ~(\ref{eq:tayloroscires}) is valid when $r_{ij}$ is sufficiently close to $d$, i.e., when $\epsilon$ is sufficiently small for $|r_{ij} - d| \le \epsilon$. Therefore, we can conclude that Eq.~(\ref{eq:tayloroscires}) describes a microvibration phenomenon in which the particles oscillate near the equilibrium point. As $\vec{F}_{ij} = - \frac{\partial \phi(r_{ij})}{\partial \vec{r}_{i}}$, the interaction force $\vec{F}_{ij}$ is obtained from Eq.~(\ref{eq:tayloroscires}) as follows:
\begin{eqnarray}
\vec{F}_{ij} = -\frac{\vec{r}_{ij}}{|\vec{r}_{ij}|} k (r_{ij} -d). \label{eq:springfrommd}
\end{eqnarray}
Equation~(\ref{eq:springfrommd}) shows the spring-force model. Mathematically, the Taylor expansion shown in Eq.~(\ref{eq:tayloroscires}) is derived under the condition $r_{ij} > d$, where the oscillator or spring is pulled away along the positive direction from the equilibrium point $d$. However, Eq.~(\ref{eq:springfrommd}) is also valid along the compressive direction because of the symmetry of $\phi$.

Recall that in the CKTG for \HL{an ideal gas}, the particles interact only when they collide, and each collision conserves energy and momentum. Moreover, the interaction duration is instantaneous and the spring force in Eq. ~(\ref{eq:springfrommd}) owing to the symmetric two-body potential is exerted only during collisions. Otherwise, the particles behave as free particles. Thus, a rigid-body contact model is more suitable than an MD model with a cutoff length. In fact, Eq.~(\ref{eq:springfrommd}) corresponds to a specific case of the DEM~\cite{Cundall1979, doi:10.1080/19648189.2008.9693050}, which is a discrete collision model based on the Kelvin--Voigt concept, where the creep effect of viscoelasticity is represented by a spring-damper system mathematically expressed as a linear second-order ordinary differential equation. In the DEM, the interaction force between the $i$th and $j$th contacting particles, $\vec{F}_{ij}$, is expressed as $\vec{F}_{ij} = -\frac{\vec{r}_{ij}}{|\vec{r}_{ij}|}(k{\theta} -\gamma \dot{{\theta}})$, where $k$ is the spring coefficient and $\gamma$ is the viscous damping coefficient. Furthermore, ${\theta}$ is the penetration depth between the $i$th and $j$th particles, and $\dot{X}$ represents the differential of $X$. Thus, Eq.~(\ref{eq:springfrommd}) is equivalent to a specific case of the DEM with $\gamma = 0$. However, because the DEM only attains the repulsive-force effect, Eq.~(\ref{eq:springfrommd}) is referenced when $r_{ij} < d$. 

\subsection{Modified kinetic theory of gas with external gravitational field}\label{sec:modkineg}
To introduce the effect of the external gravitational field, we modified the kinetic description of the particles \HL{discussed in Section~\ref{seq:CKTGexpl}}. \HL{In particular}, condition (4) for an ideal gas implies the following: The gravitational effect is offset by random collisions of particles; thus, under a sufficiently small external gravity, the particles can traverse at an approximately constant average speed across all directions by repeatedly colliding with each other. We modified condition (4) as follows to extend the CKTG \HL{for an ideal gas} to the case where the gravitational effect is not negligible. Despite the occurrence of random particle collisions, the gravitational effect remains uncanceled. Thus, the particles gradually accelerate in the direction of gravity. Briefly, the movement of a particle from wall A to wall B can be described as the throwing-down movement of volume- and size-free particles along the direction of gravity. The breakage of the isotropy of particle movements along the x-direction at wall B, while requiring the relationship in Section~\ref{seq:CKTGexpl} near wall A, is valid. To provide an appropriate explanation, we set a lower script (B) to the variables introduced in Section ~\ref{seq:CKTGexpl} when referring to the variables as a particle reaches wall B. Additionally, we set an accent when we refer to them during the traveling process between walls A and B.

According to fundamental physics, the velocity $v_{x}$ during the traveling process $\acute{v}_{x}$, is expressed as $\acute{v}_{x} = v_{x} + g t$ at time $t$. Subsequently, the velocity $v_{x}$ at the time when a particle reaches wall B, $v_{(x, B)}$, is represented by $v_{x}$ as $v_{(x, B)} = \sqrt{v_{x}^2+ 2gL}$, where $L$ is the length of one side of the cube, and $g$ represents gravity. Here, the positive direction is to the right such that the particles are accelerated from wall A to wall B.
A simple calculation using these relations yields the time interval $t_{all}$ required to perform a round trip from wall A to wall B as $t_{all} =2 \frac{-v_{x} + v_{x} \sqrt{1 + (2gL)/v_{x}^{2}}}{g}$. However, the magnitude of the potential-energy difference between walls A and B for a particle with mass $m$ is $m g L$. Hereinafter, the potential-energy difference for a particle per unit mass, $gL$, is referred to as the potential-energy density, and the kinetic energy per unit mass along the direction of gravity, $\frac{1}{2} v_{x}^{2}$, is referred to as the kinetic-energy density, unless stated otherwise. Consider the case where $gL$ is sufficiently small compared with $\frac{1}{2} v_{x}^{2}$ as $gL \ll \frac{1}{2} v_{x}^{2}$. Here, the time $t_{all}$ can be approximated as follows using the mathematical relation $ \sqrt{1+x} \simeq 1 + \frac{x}{2}$, which is valid for a sufficiently small $x$ compared with 1:
\begin{eqnarray}
t_{all} = 2\frac{-v_{x} + v_{x} \sqrt{1 + (2gL)/v_{x}^{2}}}{g} \simeq \frac{2L}{v_{x}}. 
\end{eqnarray}
Thus, if the effect of gravity is sufficiently weak, such that $gL \ll \frac{1}{2} v_{x}^{2}$, then the time required for a single particle to travel between walls A and B is similar to that in the case without gravity. Thus, the number of collisions that a single particle undergoes with a wall per unit time is similar to that in the case without gravity, i.e., $\frac{v_{x}}{2L}$. Thus, upon obtaining the impact per collision, the pressure on wall B can be derived, as in the case without gravity (Section ~\ref{seq:CKTGexpl}). The velocity of a particle along the $x$direction at the time of collision with wall B is $\sqrt{v_{x}^{2} + 2 gL}$, which becomes $-\sqrt{v_{x}^{2}+2gL}$ following the collision. The impulse $I$ is equivalent to the difference in momentum before and after the collision; therefore, we obtain $I = 2 m\sqrt{v_{x}^{2}+2gL}$. Because $gl \ll \frac{1}{2} v_{x}^{2}$, we can express $I \simeq 2 m v_{x} + \frac{2mgL}{v_{x}}$ using the approximation above, i.e., $\sqrt{1+x} \simeq 1 + \frac{x}{2}$.
Thus, the total impulse exerted on wall B by a single particle during a discrete short time $\Delta t$, $I_{tot}$ can be expressed as $I_{tot} \simeq (\frac{m v_{x}^{2}}{L} + mg) \Delta t$. We denote the force exerted by the particle on wall B in units of time as $f$. The impulse provided to a wall during $\Delta t$ is obtained as multiples of $f$ and $\Delta t$, where $I_{tot}$ = $f\Delta t$. By comparing these expressions with the expressions for $I_{tot}$, we obtain $f$ as $f \simeq \frac{m v_{x}^{2}}{L} + mg$. Accordingly, we can obtain the average force $\overline{F}$ on wall B exerted by $N$ particles after replacing the $v_{x}^{2}$ of $f$ with its average $\overline{v_{x}^{2}}$ as follows:
\begin{eqnarray}
\overline{F} \simeq N\frac{m \overline{v_{x}^{2}}}{L} + Nmg. \label{eq:fwithgravitysingle}
\end{eqnarray}
In the case without a gravitational field, $v^{2} = v_{x}^{2} + v_{y}^{2} + v_{z}^{2} = 3v_{x}^{2}$ holds because of isotropy. By contrast, when the effect of gravity is considered, the isotropy condition \HL{can be} satisfied in the vicinity of wall A. \HL{Namely}, as this is prior to the acceleration by the gravitational field, the randomness owing to the collision between the particle and wall A dominates and cancels the effect of gravity. \HL{However}, near wall B, the acceleration of particles becomes non-negligible and breaks the isotropy. The magnitude of the square of the velocity along the x direction when a particle collides with wall B is obtained as ${v_{(x, B)}^{2}} = v_{x}^{2} + 2gL$. However, as isotropy remains valid along the y- and z-directions, we have $\overline{v_{(y, B)}^{2}} = \overline{v_{(z, B)}^{2}} = \overline{v_{x}^{2}}$. Therefore, the mean square of the velocity at wall B, $\overline{v_{B}^{2}}$, is expressed as $\overline {v_{B}^{2}} = 3\overline{v_{x}^{2}} + 2gL = \overline{v^{2}} + 2gL$.

Following a simple calculation using these relationships pertaining to the mean square of the velocity, the pressure $P$ on wall B can be obtained from Eq.~(\ref{eq:fwithgravitysingle}) using the average velocity $\overline{v^{2}}$, the relationship between the pressure $P$ and $\overline{F}$ as $P=\overline{F}/L^{2}$, and $V=L^{3}$.
\begin{eqnarray}
PV \simeq \frac{Nm\overline{v^{2}}}{3} + NmgL. \label{eq:modifEoS} 
\end{eqnarray}
Based on the discussion in Section~\ref{seq:CKTGexpl}, the first term on the right-hand side of Eq.~(\ref{eq:modifEoS}) is equal to $Nk_BT$. Therefore, Eq.~(\ref{eq:modifEoS}) can be written as 
\begin{eqnarray}
P \simeq \frac{1}{V} \biggl[ Nk_{B} T + NmgL \biggr]. 
\end{eqnarray}
Consequently, the following pressure expression can be obtained after replacing the first term in the parentheses on the right-hand side of Eq.~(\ref{eq:prestwobodypot}), $N k_{B} T$, i.e., the contribution of the EOS to an ideal gas with $N k_{B} T + N mgL$:
\begin{eqnarray}
\braket{\mathcal{P}} = \frac{1}{V} \biggl[ N k_{B} T + N mgL - \frac{1}{3}\sum _{i=1} \sum _{j<i} r_{ij} \frac{d \phi(r_{ij})}{d r_{ij}} \biggr]. \label{eq:modifiedLocPFin} \nonumber \\
\end{eqnarray}
Equation~(\ref{eq:modifiedLocPFin}) represents the generic form of the average pressure for $N$-particle interacting systems with an external gravitational field.
The second term in the parentheses on the right-hand side implies that the pressure at a distance $L$ from the position of the base point along the direction of the gravitational field is proportional to the distance $L$, which exceeds the pressure at the base point under the influence of the gravitational field. 

Alternatively, Eq.~(\ref{eq:modifiedLocPFin}) can be obtained as follows: Equation~(\ref{eq:presone}) is derived from the equation of motion for the $i$th particle in the absence of a gravitational field, i.e., $m \ddot{\vec{r}}_{i} = \vec{F}_{int}^{(i)} + \vec{F}_{wall}^{(i)}$. Consequently, the following equation of motion around wall B for each particle can be assumed: $m \ddot{\vec{r}}_{B} = \vec{F}_{int} + \vec{F}_{wall}$, where $\vec{r}_{B}$ is the position of the particle. Additionally, $v_{B}$ is the derivative of $r_{B}$ with respect to time. Accordingly, similar to the case wherein Eq.~(\ref{eq:presone}) is derived from $m \ddot{\vec{r}}_{i} = \vec{F}_{int}^{(i)} + \vec{F}_{wall}^{(i)}$, we can derive Eq.~(\ref{eq:modifiedLocPFin}) from $m \ddot{\vec{r}}_{B} = \vec{F}_{int} + \vec{F}_{wall}$ after applying the relationship $\overline{v_{B}^{2}} = \overline{v^{2}} + 2gL$.

We demonstrated our \HL{prediction} regarding the asymmetry of pressure in the presence of an external gravitational field, \HL{that is}, in N-particle interaction systems, where energy and momentum are conserved before and after collisions, the downwind pressure of gravity exceeds the upwind pressure, provided that the gravitational effect is not negligible. In general, this nature is evident for continua. However, in molecular kinetic systems with a large mean-free path relative to the representative length (large Kn), the gravitational effect is typically considered negligible because it is damped by the randomness of the collisions between the molecules. \HL{This poses the question of whether the abovementioned holds for highly rarefied gases.} We will show that even if the potential-energy density is approximately 1 $\%$ of the kinetic-energy density, the gravitational effect remains uncanceled within the randomness of collisions between particles. Furthermore, the pressure on the wall placed downwind along the direction of gravity exceeds that on the wall placed upwind.

\begin{figure*}[t]
\vspace{-61.0cm}
\hspace{-1.0cm}
\includegraphics[width=4.6\textwidth, clip, bb= 0 0 5295 5021]{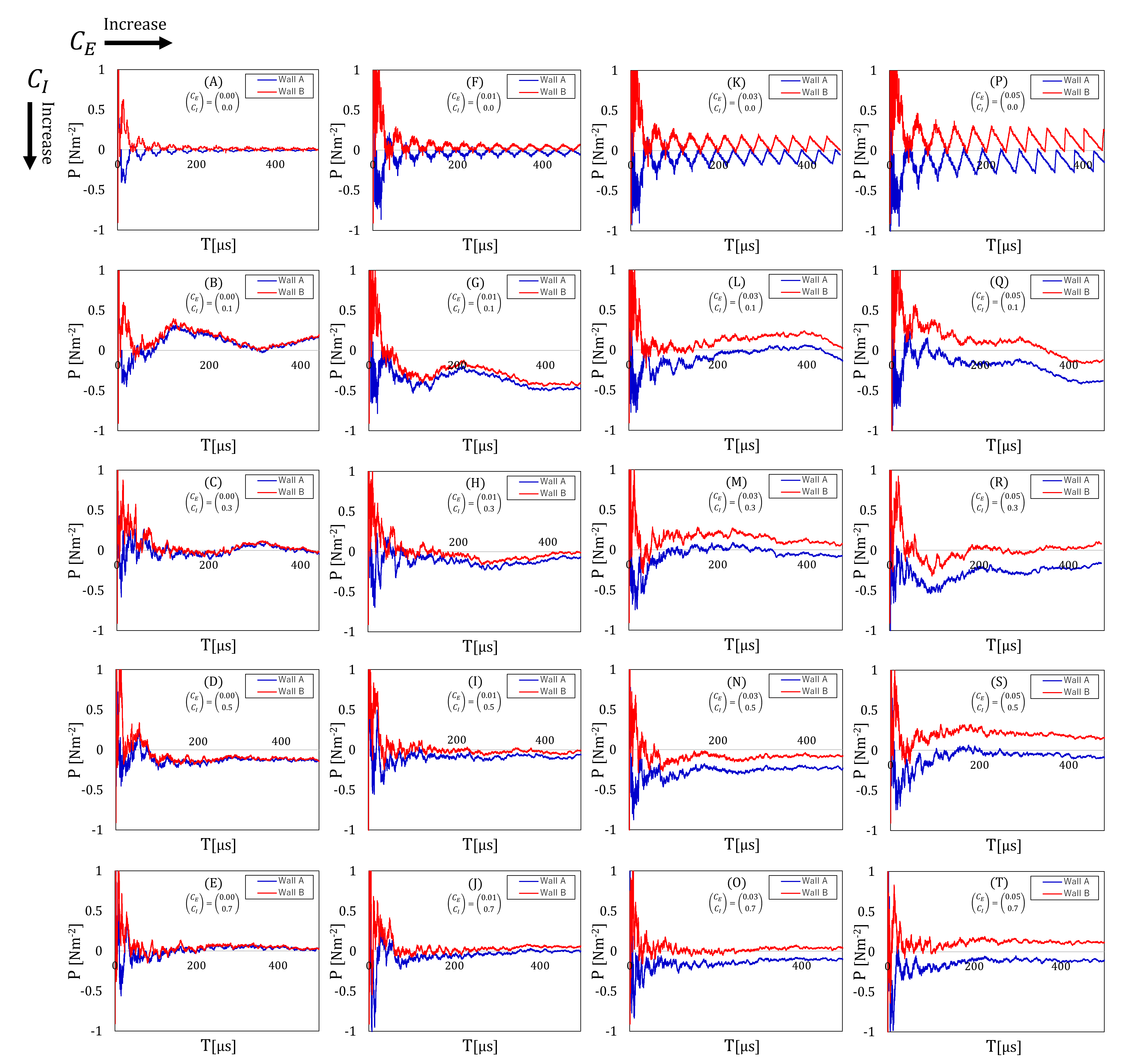}
\caption{Time-series plots of pressure values on walls A and B for different values of $C_E$ (ratio of potential-energy density to kinetic-energy density) and $C_I$ (square root of collision interaction probability) within 0.01--0.05 and 0.1--0.7, respectively. Values of $(C_E, C_I)$ for each simulation are shown at top center of each subfigure for reference. Each $C_E$ and $C_I$ in the plots increases with the column or row, respectively.}
\label{fig:Figure2}
\end{figure*}

\begin{figure*}[t]
\vspace{-77.5cm}
\hspace{+2.0cm}
\includegraphics[width=4.4\textwidth, clip, bb= 0 0 3874 4826]{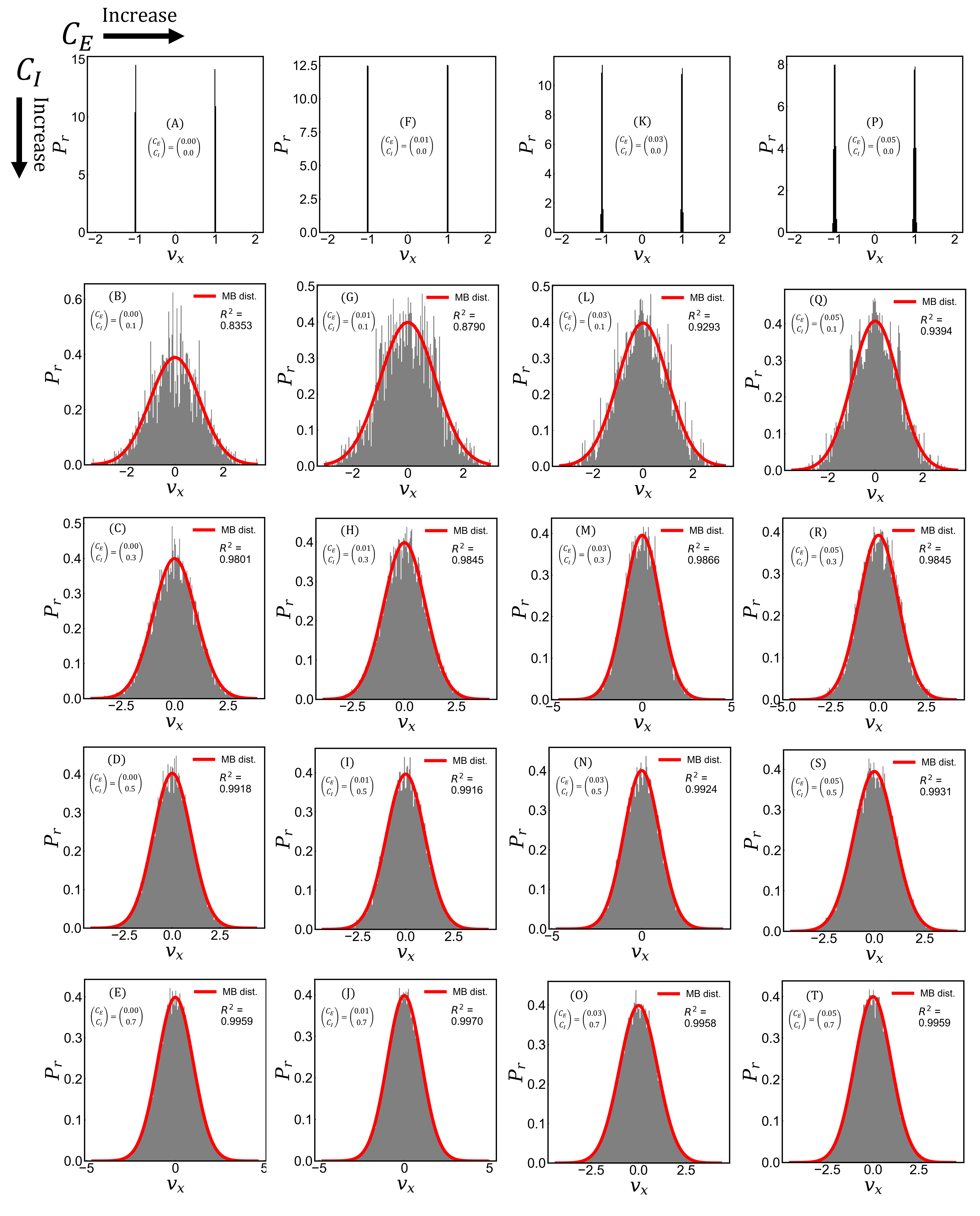}
\caption{Normalized histograms of velocity distributions, i.e., probability density functions (PDF) of velocity profiles in respective cases presented in Fig.~\ref{fig:Figure2} obtained by considering arithmetic mean of spatial distribution between 250 and 500 $\mu s$.}
\label{fig:Figure3}
\end{figure*}

\begin{figure*}[t]
\vspace{-25.5cm}
\hspace{+1.2cm}
\includegraphics[width=4.2\textwidth, clip, bb= 0 0 4446 2089]{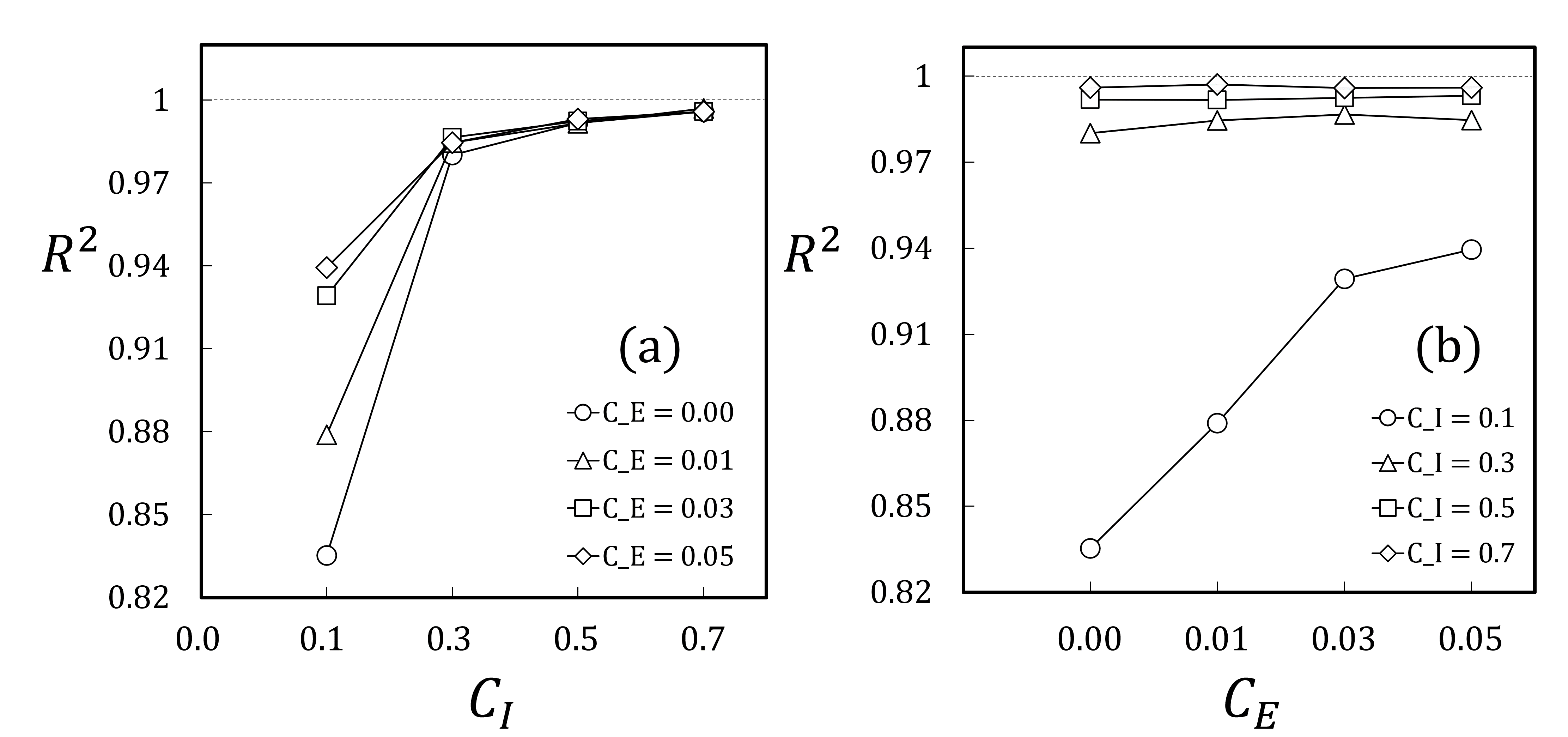}
\caption{Dependence of $R^2$ on parameters $C_{E}$ and $C_{I}$ for all cases investigated.}
\label{fig:Figure4}
\end{figure*}

\section{Numerical analysis} \label{sec:NumAnalysis}
\HL{We} performed 2D particle-collision simulations using the DEM as follows: First, the computational domain was set as a square domain with of side $L$. Furthermore, gravity $g$ was added along the $x$-axis direction, which was positive from walls A to B. Each particle retained the following physical properties: mass $m$; radius $r$; coordinate $p$; velocity $v$; acceleration $a$; applied force $f$; kinetic energy $E_{k}$; potential energy $E_{p}$; array of time-accumulated impulses provided by itself to each of the four walls, ${\rm I}[w]$; and accumulated counts of collisions with each wall, ${\rm C}[w]$, where $w$ is a serial number from 1 to 4, each of which was set to walls A, B, C, and D, respectively. Walls A and B faced each other, as were walls C and D. Here, $p$, $v$, $a$, and $f$ are 2D vectors. In the interaction between particles, the repulsive force in Eq.~(\ref{eq:springfrommd}) was calculated between two particles when they were in contact with each other. The spring constant was determined based on the Hertz--Mindlin contact theory, as in ordinary DEM calculations. The discrete small time is denoted by $dt$. The velocity Verlet method~\cite{PhysRev.159.98} was used to update the physical properties of the particles following the collision calculations for each discrete time. The average total force $f$ per unit time exerted by the particles on wall $w$ was calculated from ${\rm I}[w]$ and ${\rm C}[w]$. The average pressure $P$ at each wall was updated by dividing the force by the length $L$ and physical time at each time step of the simulation. Although dividing the force by $L$ to obtain $P$ is valid for 2D cases, for clarity, the motion along the depth direction in the 3D case was interpreted as being projected onto the xy plane, or the depth direction was assumed to have a unit length of thickness. Furthermore, the unit of pressure in the simulation results presented in the next section is $[N/m^{2}]$.

In the initial state, $N$ particles were arranged with numbers $N_{x}$ and $N_{y}$ along each direction of the computational domain (thus, $N = N_{x} \times N_{y}$). We define $C_{E}$ as the ratio of the potential-energy density to the kinetic-energy density. Recall that the kinetic-energy density is $\frac{1}{2}v^{2}$ and the potential-energy density is $gl$. Thus, we obtain the relationship $g=C_{E}\frac{v_{init}^{2}}{2L}$, \HL{where $v_{\rm init}$ represents the initial velocity of each particle}. We determined $g$ based on this relationship. Furthermore, we introduced the parameter $C_{I}$ for the interaction probability. In the following, we refer to the $i$th and $j$th particles as particles $i$ and $j$, respectively. In the simulations, when particle $j$ entered the vicinity of particle $i$ (an area twice the size of its diameter along each direction wherein the particle inside it can collide with particle $i$ in the near future time step), we randomly determined (probability $C_{I}$) whether particles $i$ and $j$ would interact. Here, particles $i$ and $j$ ``rolled the dice'' separately, and if both were true, then the two particles would be registered in the list of pair particles. Therefore, we obtained the relationship $p = C_{I}^2$, where $p$ indicates the probability of particles $i$ and $j$ colliding with each other. \HL{Figure~\ref{fig:Figure1}(b) shows a schematic illustration of the stochastic decision-making process for particles $i$ and $j$ regarding whether to collide with one another.} If the judgement is true, then the interaction force between particles $i$ and $j$ is calculated, where particles $i$ and $j$ are in contact. However, if the particles do not collide and particle $j$ remains in the vicinity of particle $i$, then the relationship between particles $i$ and $j$ is re-set. As with the parameter $C_{E}$, simulations were performed under different ranges of $C_I$. We considered $L$ to be the representative length. To render the simulation results applicable to a wide range of problems, we first expressed the remaining parameters as functions of $L$. We defined $C_{r}$ as the ratio of the particle radius $r$ to the representative length $L$. Thus, $r = C_{r} L$. The parameter $C_{r}$ was set to 1/100 in our simulations.

To ensure numerical stability, we allowed a penetration depth of up to 1/100 of the radius between two particles. Additionally, we set $dt = 1.0 \times 10^{-5}$, assuming that the penetration depth was computed in approximately 100 increments along each compression and stretching direction. This is primarily because the computation of a period in 5--50 increments in the DEM is numerically stable~\cite{DANBY2013211, HART1988117, OTSUBO201767, PhysRevLett.67.1751, refId01997, LANGSTON1995967}. We prepared a sufficient number of discrete time steps for each collision and computed $5 \times 10^{7}$ time steps for 500 $\mu s$ in physical time. Consequently, the relative error of the total energy from the initial energy remained within 0.0005 $\%$ after all steps of the simulation were completed when $C_{I} = 0.0$. As mentioned above, the units along the time direction were set to $\mu s$. The $x$-component $v_{x}$ and the $y$-component $v_{y}$ of the velocity of each particle were randomly set to either -1/20$L$ or 1/20$L$ $\mu s^{-1}$. The length $L$ was assumed to be a dimensionless quantity and set to 20 in all cases. Additionally, we set $N_{x}$ and $N_{y}$ to 20, and $N$ to 400 particles. Thus, our system is characterized by a set of variables $(L, N, C_r, C_{E}, C_{I}, dt)$. Among these, we set $(L, N, C_{r}, dt)$ = (20, 400, 0.01, $1.0\times 10^{-5}$) for all the calculations. The reasons for setting $(L, N)$ = (20, 400) are elaborated in the Discussion section. We performed simulations for different pairs of the ratio of the potential-energy density to the kinetic-energy density $C_{E}$ and for different frequencies of particle collisions $C_{I}$.
\HL{We provide two example videos for cases $(C_{E}, C_{I})$ = (0.03, 0.0) and $(C_{E}, C_{I})$ = (0.05, 0.5) herein as supplementary data. Each video is approximately 1 min long, which corresponds to 2 s in physical time. The video for the former conditional case is {\tt Video\_CE\_003\_CI\_00.mp4}, which shows the case without particle interactions, and confirmed that all particles penetrated each other while interacting only with the walls. By contrast, the video for the latter case is {\tt Video\_CE\_005\_CI\_05.mp4}, which shows the case with particle interactions at a rate of $p = C_{I}^{2} = 0.25$ and confirmed that some particles collided with each other while others penetrated each other. Additionally, the latter exemplifies the case of an external gravitational field that generates potential energy equivalent to 5$\%$ of the total kinetic energy, the magnitude of which is the maximum case for the tests. Clearly, the strongness of the gravitational field does not cause significantly affect the distribution in the direction of gravity. The corresponding videos for our preprinted manuscript are available at {\tt https://arxiv.org/abs/2409.00454}.}

Figure~\ref{fig:Figure2} shows the time-series plots of the pressure values on walls A and B for different values of $C_E$ and $C_I$, which changed within 0.01--0.05 and 0.1--0.7, respectively. The values of $(C_E, C_I)$ for each simulation are shown at the top center of each subfigure for reference. In Fig.~\ref{fig:Figure2}, each $C_E$ and $C_I$ in the subfigures increased with the column and row, respectively. In particular, the top-leftmost subfigure in Fig.~\ref{fig:Figure2}(A) shows the case of $(C_E, C_I)$ = $(0.0, 0.0)$, which featured no external gravitational field and no collisions between the particles. The important observations were as follows: First, Fig.~\ref{fig:Figure2}(A) shows the feature of the CKTG for an ideal gas explained in Section ~\ref{seq:CKTGexpl}, that is, the pressure varies periodically along the time direction. This is primarily because each particle can traverse between walls A and B with the same constant velocity as the initial velocity owing to the absence of collision among the particles. The periods of pressure variation on walls A and B were opposite and their averages were maintained at approximately zero. Second, based on (A), (F), (K), and (P) in Fig.~\ref{fig:Figure2}, the pressure at wall B exceeded that at wall A as $C_E$ increased, i.e., the ratio of the potential-energy density to the kinetic-energy density increased, while $C_I$, the square root of the collision interaction probability, remained zero. In fact, the pressure changed periodically, similar to (A), because the system remained deterministic owing to the same principle of $C_I$ = 0, whereas the acceleration of the particles due to the increase in the magnitude of gravity contributed to the pressure at wall B compared to (A).

Third, after the second row in Fig.~\ref{fig:Figure2}, the parameter for the frequency of particle collisions $C_I$ became nonzero. Consequently, randomness due to particle collisions appeared, and periodic pressure changes were no longer observed. By focusing on the $C_E$ = 0 series at the left end of Fig.~\ref{fig:Figure2}, the time dependence of the pressure stabilized as $C_I$ increased, particularly when approaching the steady state for $C_I$ = 0.5 and 0.7, as shown in (D) and (E) of Fig.~\ref{fig:Figure2}. Conversely, the time dependence of the pressure remained unstable when $C_I$ = 0.1, as shown in (B). The results of (D) and (E) can be explained as follows: In this system, the randomness of the particle motion increases rapidly with $C_I$. Specifically, 49 $\%$ of the particles collided on average when $C_I$ = 0.7, where $C_I^{2}$ represents the collision probability. The increase in randomness caused the system to transition from being deterministic, as observed in (A), to being statistical, followed by adherence to equilibrium statistical mechanics. By contrast, (B) is a transient phase from (A) to (C)--(E) and therefore did not reach a steady state. Interestingly, the same discussion can be applied to the remaining cases with $C_E$ for 0.01--0.05, i.e., the cases with non-negligible gravity. This phenomenon occurred independently of gravitational effects.

Figure~\ref{fig:Figure3} shows the normalized histograms of the velocity distributions, i.e., the probability density functions of the velocity profiles in the respective cases presented in Fig.~\ref{fig:Figure2} obtained by considering the arithmetic mean of the spatial distribution between 250 and 500 $\mu s$. The results shown in each figure correspond to those shown in Fig.~\ref{fig:Figure2} under the same alphabetical subscripts (A)–(T). The distributions are shown in gray. The red curves in each subfigure represent the fitting results using the MB distribution, which is expressed as $f(v_{x}) = \sqrt{\frac{m}{2 \pi k_{B} T}} {\rm exp}[ -\frac{m}{2k_B T} v_x^{2}]$ using the least mean square. These fitting results corroborated the third observation, i.e., the velocity distributions exhibited better consistency with the MB distribution with an accuracy of $R^{2}$ = 0.99, or were better in all cases of $C_I$ = 0.5 and 0.7. Here, $R^{2}$ is the coefficient of determination. In the cases of $C_I$ = 0.5 and 0.7, randomness due to particle collisions dominated in the entire dynamics, and the system adhered to equilibrium statistical mechanics. By contrast, at $C_I$ = 0.3, regardless of $C_{E}$, $R^{2}$ deteriorated and became around 0.98, \HL{and it} deteriorated significantly at $C_{I}$ = 0.1, where \HL{we can see that} the system deviated from the thermodynamic equilibrium state. Finally, in the case of $C_I$ = 0, each particle maintained the initial velocity profiles (which was provided as either -1 or 1) during the simulations. Thus, the system dynamics can be considered to be deterministic in this case.

Figure~\ref{fig:Figure4} depicts the correlation between $R^2$ and each of parameters $C_E$ and $C_I$ for all cases. Evidently, $R^2$ increased and approached 1 as $C_I$ increased (Fig.~\ref{fig:Figure4}(a)). The results show that as the randomness due to particle collisions increased, the distribution approached an MB distribution, thus indicating that the system began to obey equilibrium statistical mechanics. The same was observed for the range $C_E > 0$, where the pressure difference between the walls was significant. This indicates that when the potential energy was relatively small compared with the kinetic energy, at 1 $\%$–5 $\%$, a pressure difference between the walls emerged while the characteristics of equilibrium statistical mechanics, as shown by the MB distribution, were maintained. When $C_I$ = 0.1, at which point the system was in a transitional state between deterministic and statistical mechanics, a greater gravitational force  stabilized the system more effectively and the MB distribution was realized. This is confirmed by the line plots with circular symbols in Fig.~\ref{fig:Figure4}(b), which shows the results for the case where $C_I$ = 0.1.

\section{Discussion}\label{seq:discuss}
The findings of this study are as follows: (a) When the potential-energy density was sufficiently small (but not negligible) compared with the kinetic-energy density, the gas pressure increased in proportion to the distance from the gravitational reference point. (b) In the absence of interactions between the particles, the system exhibited deterministic behavior. When the interaction parameter was $C_I$ = 0.1 or higher, i.e., the probability of particle collision was 0.01 $\%$ or higher, the system changed from deterministic to statistical and then exhibited equilibrium statistical mechanics when the probability of collision was 0.25 $\%$ or higher. (c) In the transition region from the deterministic to the statistical system state (in the case of collision probability = 0.01 owing to $C_I$ = 0.1), the system exhibited unstable behavior. (d) The results in (b) and (c) were independent of the external gravitational field. Thus, even when the potential energy was relatively small compared with the kinetic energy, at 1 $\%$–5 $\%$, a pressure difference emerged between the walls while the characteristics of equilibrium statistical mechanics, as shown by the MB distribution, were maintained. (e) Even during the transition from a deterministic to a statistical state, where the particle collision probability was 0.01 $\%$, the addition of an external gravitational field stabilized the system. Moreover, the velocity distribution in the direction of the gravitational field approached that of the system described by the MB distribution.

\HL{The essential mechanism of the transition from deterministic to stochastic described in (b) can be understood as follows. To make it easier to understand, we refer to the two videos provided in the supplementary data: (i) { \tt Video\_CE\_003\_CI\_00.mp4} for the simulation without particle collisions, and (ii) { \tt Video\_CE\_005\_CI\_05.mp4} for the simulation in which some particles collided with each other while others penetrated each other. As mentioned earlier, the average pressure that a wall receives from the particles at a certain point in time was obtained by dividing the accumulated value of the impulses applied to the wall by the elapsed physical time at that point. This operation corresponds to obtaining the ensemble average of the pressure in the time and spatial directions and is based on the same concept used to obtain the pressure in the CKTG for an ideal gas, as presented in Section~\ref{seq:CKTGexpl}. In the simulation with $C_{I} = 0.0$ in (i), the initial velocity of the particles in each spatial direction was randomly set to -1 or 1. However, because the particles did not interact with each other in this case, they traversed in a straight line and maintained their initial velocity and direction until they collided with one of the four walls. After colliding with the wall, the direction and sign of the velocity reversed with respect to the angle of incidence (see $\psi$ in Fig.~\ref{fig:Figure1}(a)). Thus, each particle repeated a fixed pattern of movement, e.g., $A \rightarrow C \rightarrow B \rightarrow D$ for particle X, where the behavior pattern of each particle is determined by the initial conditions and remains unchanged throughout the simulation; hence, it is known as deterministic. When viewed from the wall, a fixed number of particles determined in the initial state approached the wall, collided with it, and then departed from it---this process was repeated. As the particles departed from the wall, the total accumulated number of collisions or impulses remained unchanged---only the physical time elapsed. Accordingly, the average pressure, as calculated by the ratio of these two, decreased. However, when the set of particles reapproached the wall, the opposite occurred and the average pressure increased. Therefore, the pressure on each wall was periodic at $C_{I} = 0.0$. }

\HL{Meanwhile, if $C_{I}$ is nonzero, then each pair of particles collides with probability $p = C_{I}^{2}$ when they enter the vicinity of each other, as shown in Fig.~\ref{fig:Figure1}(b).
{ \tt Video\_CE\_005\_CI\_05.mp4} presents an example where $C_{I} = 0.5$. The pairs of particles that can collide varied in each trial. Accordingly, the next destination of each particle directed by each collision event was randomly changed in the pair-selection trial as the particles approached each other. When viewed from the wall, one cannot guarantee that every particle that collides with a wall will collide with it again because the collision pairs are selected randomly. In summary, the system becomes stochastic for nonzero $C_{I}$ values. Notably, our simulation results confirmed that even when a weak gravitational field is applied to such a stochastic system, it still obeys the MB distribution, i.e., the local equilibrium distribution function in classical statistical mechanics.}

Meanwhile, observation (d) can be explained as follows: Under a gravitational field, we can replace $v^{2}$ by $v^{2} + 2gL$ in the MB distribution as $f(v_{x}) = \sqrt{\frac{m}{2 \pi k_{B} T}} {\rm exp}[ -\frac{m}{2k_B T} (v_x^{2}+2gL)]$, which can be rewritten as $f(v_{x}) = C\sqrt{\frac{m}{2 \pi k_{B} T}} {\rm exp}[ -\frac{m}{2k_B T} v_x^{2}]$, where $C$ = ${\rm exp}[-\frac{mgL}{k_{B} T}]$. That is, each factor in the MB distribution is multiplied by $C$, which represents the contribution of the potential energy ($mgL$) of each particle. This does not affect the resulting probability distribution because $C$ is constant. In this case, the system retains the properties of equilibrium statistical mechanics.

Finally, to obtain insights into the physical properties, we discuss the mean-free path and RMS velocity of the target system for length $L$ = 20. Consider the case $(C_I, C_E)$ = (0.0, 0.00) in Figs.~\ref{fig:Figure2}(A) and ~\ref{fig:Figure3}(A), where interactions between the particles and the influence of an external gravitational field are absent. Here, the particles traversed 1/20 of the length of one side $L$ in 1 $\mu s$. This corresponds to 25,000 round trips of the domain in 1 s. For comparison, the RMS velocity of a helium molecule at 0 ${}^\circ$C is approximately 1304 m/s, as calculated using $v = \sqrt{3 k_B T/m}$. Therefore, if a helium atom at 0 ${}^\circ$C is enclosed in a cube measuring 1 inch = 2.54 cm = 0.0254 m on each side, then it would make approximately 25669 round trips per second, on average. Thus, the molecular velocity for $L$ = 20 is comparable in scale to the RMS velocity of helium gas at 0 ${}^\circ$C, provided that we regard it as an ideal gas.

The next step involves estimating the mean-free path. Generally, the mean-free path $l$ in the three-dimensional case is obtained as follows: First, we consider a particle as a sphere of radius $r$. If a particle traverses for a duration of $t$ s at an average velocity of $v$, then the volume of space occupied by the particle at that time is expressed as $Svt$. Here, $S$ denotes the cross-sectional area of a sphere with radius $2 r$. We denote the number density of the particles as $n$. The number of collisions is $Svtn$ if all particles, except themselves, are at rest. However, if the particles are traversing relatively, then $v$ is replaced by their relative velocity, which is $\sqrt{2}$ times the average velocity $v$. Thus, the number of collisions is $\sqrt{2Svtn}$. Dividing the average distance traversed by a particle $vt$ by the number of collisions $\sqrt{2Svtn}$ yields $l=(\sqrt{2} S n)^{-1}$. In a 2D system, the volume of space is perceived as an area; thus, we can perceive the cross-sectional area $S$ as the diameter $D$ of a sphere with radius $2r$. Therefore, $S=4r$, which yields $l=(4\sqrt{2}rn)^{-1}$, This is perhaps the most primitive estimate of the mean-free path. Because the particle number density is $400/L^{2}$ and the radius is $r=L/100$, $l$ is expressed as $l=L/16\sqrt{2}$. If $L$ is 1 inch = 2.54 cm = 0.0254 m, as in the example above, then $l \approx 0.11225$ cm. The mean-free path of a helium atom at 0${}^\circ$C and 1 Torr (=101325/760 Pa) is approximately 0.0139 cm when the atomic radius of helium is estimated to be $1.07 \times 10^{-10}$ m~\cite{jccs201500046}; therefore, the mean-free path length in this analysis is approximately 8.0 times longer than that of helium gas at 0${}^\circ$C and 1 Torr. However, another study estimated the atomic radius of helium to be $0.3023 \times 10^{-10}$ m~\cite{Harbola_2011, ZHU201428}, which yielded $l = 0.17417$ cm. The mean-free path of our test was approximately 0.64 times longer than that of the minimum estimate. Thus, the mean-free path in our test was within a realistic helium range at 0${}^\circ$C and 1 Torr. We compared the RMS velocity and mean-free path to those of helium gas at 0${}^\circ$C as a reference for the application of our approach to an actual gas system in future investigations.

\section{Conclusion}
\HL{In this study, we examined the effects of an external gravitational field on highly rarefied gases in the transition regime near free molecular flows. In our theoretical study}, we rederived the classical kinetic theory for an ideal gas in terms of the kinetics of the constituent particles to account for the effect of accelerating the particles by an external gravitational field. Subsequently, we derived an extended expression for the VPE \HL{as a generic description of the dynamics under an external gravitational field in this regime}. \HL{We used the soft-sphere model for the following reasons: In highly rarefied gases, short-range and instantaneous collisional interactions are dominant. Thus, we expanded the asymmetric two-body potential in the VPE and retained only the contribution of the short-range interaction, thus achieving a soft-sphere model that represents the interaction in the collision direction as a harmonic oscillation.}
In our collision simulations, we defined two parameters. The first parameter represents the collision probability between each pair of approaching particles, and the second represents the ratio of the magnitude of the external potential energy to the total kinetic energy of the particles. The behavior of the system was analyzed by varying the values of these two parameters. 

\HL{Our analysis proved that if the potential energy due to the external gravitational field is sufficiently small (but not negligible), i.e., 1 $\%$--5 $\%$ of the total kinetic energy, then a pressure difference would emerge between the walls while the properties of equilibrium statistical mechanics, as shown by the MB distribution, were maintained. The remaining important observations can be summarized as follows: In the absence of particle interactions, the system exhibited deterministic behavior. However, as the probability of particle collisions increased, the system changed from deterministic to statistical, followed by adherence to equilibrium statistical mechanics. During the transition from the deterministic to the statistical state, the system exhibited unstable behavior; these observations were independent of the external potential field. More importantly, an external gravitational field did not determine whether the system would adhere to equilibrium statistical mechanics. Instead, during the transition from a deterministic to a statistical state, the addition of an external gravitational field stabilized the system and rendered it more similar to the system described by the MB distribution. In summary, highly rarefied gases obey the MB distribution even when placed under weak gravitational fields.}







\section*{Data availability statement}
This manuscript contains no relevant data.

\section*{Acknowledgment}
This study was supported by JSPS KAKENHI (grant number 22K14177) and JST PRESTO (grant number JPMJPR23O7).
The authors thank Editage (www.editage.jp) for English language editing.
The author would also like to express his gratitude to his family for their moral support and encouragement.

\bibliographystyle{h-physrev3}
\bibliography{reference}

\begin{thebibliography}{10}

\bibitem{laurendeau2005statistical}
N.~Laurendeau,
\newblock {\em Statistical Thermodynamics: Fundamentals and Applications}
  (Cambridge University Press, 2005).

\bibitem{GUO20151}
Y.~Guo and M.~Wang,
\newblock Phonon hydrodynamics and its applications in nanoscale heat
  transport, Physics Reports {\bf 595}, 1 (2015),
\newblock Phonon hydrodynamics and its applications in nanoscale heat
  transport.

\bibitem{ESPOSITO20051}
R.~Esposito and M.~Pulvirenti,
\newblock Chapter 1 - from particles to fluids,
\newblock volume~3 of {\em Handbook of Mathematical Fluid Dynamics}, pp. 1--82,
  North-Holland, 2005.

\bibitem{ARUMUGAPERUMAL2015955}
D.~{Arumuga Perumal} and A.~K. Dass,
\newblock A review on the development of lattice boltzmann computation of macro
  fluid flows and heat transfer, Alexandria Engineering Journal {\bf 54}, 955
  (2015).

\bibitem{doi:10.1098/rsta.1916.0006}
S.~Chapman and J.~Larmor,
\newblock Vi. on the law of distribution of molecular velocities, and on the
  theory of viscosity and thermal conduction, in a non-uniform simple monatomic
  gas, Philosophical Transactions of the Royal Society of London. Series A,
  Containing Papers of a Mathematical or Physical Character {\bf 216}, 279
  (1916).

\bibitem{chapman1990mathematical}
S.~Chapman and T.~G. Cowling,
\newblock {\em The mathematical theory of non-uniform gases: an account of the
  kinetic theory of viscosity, thermal conduction and diffusion in gases}
  (Cambridge university press, 1990).

\bibitem{Cercignani2002}
C.~Cercignani and G.~M. Kremer,
\newblock {\em Chapman-Enskog Method} (Birkh{\"a}user Basel, Basel, 2002), pp.
  111--147.

\bibitem{RAPP2017243}
B.~E. Rapp,
\newblock Chapter 9 - fluids,
\newblock in {\em Microfluidics: Modelling, Mechanics and Mathematics}, edited
  by B.~E. Rapp, Micro and Nano Technologies, pp. 243--263, Elsevier, Oxford,
  2017.

\bibitem{BOYD2005669}
I.~D. Boyd,
\newblock Numerical modeling of spacecraft electric propulsion thrusters,
  Progress in Aerospace Sciences {\bf 41}, 669 (2005).

\bibitem{10.1119/1.10664}
S.~Hjalmars,
\newblock Evidence for boltzmann'h as a capital eta, American Journal of
  Physics {\bf 45}, 214 (1977).

\bibitem{PhysRev.94.511}
P.~L. Bhatnagar, E.~P. Gross, and M.~Krook,
\newblock A model for collision processes in gases. i. small amplitude
  processes in charged and neutral one-component systems, Phys. Rev. {\bf 94},
  511 (1954).

\bibitem{Butta2021}
P.~Butt{\`a}, M.~Hauray, and M.~Pulvirenti,
\newblock Particle approximation of the bgk equation, Archive for Rational
  Mechanics and Analysis {\bf 240}, 785 (2021).

\bibitem{https://doi.org/10.1002/anie.199009921}
W.~F. van Gunsteren and H.~J.~C. Berendsen,
\newblock Computer simulation of molecular dynamics: Methodology, applications,
  and perspectives in chemistry, Angewandte Chemie International Edition in
  English {\bf 29}, 992 (1990).

\bibitem{Nose10061984}
S.~Nose,
\newblock A molecular dynamics method for simulations in the canonical
  ensemble, Molecular Physics {\bf 52}, 255 (1984).

\bibitem{HANSSON2002190}
T.~Hansson, C.~Oostenbrink, and W.~{van Gunsteren},
\newblock Molecular dynamics simulations, Current Opinion in Structural Biology
  {\bf 12}, 190 (2002).

\bibitem{Grad1958}
H.~Grad,
\newblock {\em Principles of the Kinetic Theory of Gases} (Springer Berlin
  Heidelberg, Berlin, Heidelberg, 1958), pp. 205--294.

\bibitem{Cercignani01011972}


\bibitem{Lanford1975}
O.~E. Lanford,
\newblock {\em Time evolution of large classical systems} (Springer Berlin
  Heidelberg, Berlin, Heidelberg, 1975), pp. 1--111.

\bibitem{cercignani2012many}
C.~Cercignani, U.~Gerasimenko, and D.~Y. Petrina,
\newblock {\em Many-particle dynamics and kinetic equations}volume 420
  (Springer Science \& Business Media, 2012).

\bibitem{10.1093/imamat/hxr004}
H.~Struchtrup and P.~Taheri,
\newblock Macroscopic transport models for rarefied gas flows: a brief review,
  IMA Journal of Applied Mathematics {\bf 76}, 672 (2011).

\bibitem{10.1063/1.2363381}
E.~de~Miguel and G.~Jackson,
\newblock {The nature of the calculation of the pressure in molecular
  simulations of continuous models from volume perturbations}, The Journal of
  Chemical Physics {\bf 125}, 164109 (2006).

\bibitem{allen2017computer}
M.~P. Allen and D.~J. Tildesley,
\newblock {\em Computer simulation of liquids} (Oxford university press, 2017).

\bibitem{doi:10.1098/rspa.1924.0081}
J.~E. Jones and S.~Chapman,
\newblock On the determination of molecular fields. from the variation of the
  viscosity of a gas with temperature, Proceedings of the Royal Society of
  London. Series A, Containing Papers of a Mathematical and Physical Character
  {\bf 106}, 441 (1924).

\bibitem{doi:10.1098/rspa.1925.0147}
J.~E. Lennard-Jones and S.~Chapman,
\newblock On the forces between atoms and ions, Proceedings of the Royal
  Society of London. Series A, Containing Papers of a Mathematical and Physical
  Character {\bf 109}, 584 (1925).

\bibitem{PhysRevA.2.221}
J.-P. Hansen,
\newblock Phase transition of the lennard-jones system. ii. high-temperature
  limit, Phys. Rev. A {\bf 2}, 221 (1970).

\bibitem{fleagle1981introduction}
R.~G. Fleagle and J.~A. Businger,
\newblock {\em An introduction to atmospheric physics} (Academic Press, 1981).

\bibitem{merriam1992atmospheric}
J.~Merriam,
\newblock Atmospheric pressure and gravity, Geophysical Journal International
  {\bf 109}, 488 (1992).

\bibitem{Ghiroldi_Gibelli_2015}
G.~P. Ghiroldi and L.~Gibelli,
\newblock A finite-difference lattice boltzmann approach for gas microflows,
  Communications in Computational Physics {\bf 17}, 1007^^e2^^80^^931018
  (2015).

\bibitem{YUAN201625}
Y.~Yuan and S.~Rahman,
\newblock Extended application of lattice boltzmann method to rarefied gas flow
  in micro-channels, Physica A: Statistical Mechanics and its Applications {\bf
  463}, 25 (2016).

\bibitem{10.1115/1.4031000}
M.~Watari,
\newblock Is the lattice boltzmann method applicable to rarefied gas flows?
  comprehensive evaluation of the higher-order models, Journal of Fluids
  Engineering {\bf 138}, 011202 (2015).

\bibitem{Kruger2017}
T.~Kr{\"u}ger {\em et~al.},
\newblock {\em The Lattice Boltzmann Equation} (Springer International
  Publishing, Cham, 2017), pp. 61--104.

\bibitem{PhysRevE.56.6811}
X.~He and L.-S. Luo,
\newblock Theory of the lattice boltzmann method: From the boltzmann equation
  to the lattice boltzmann equation, Phys. Rev. E {\bf 56}, 6811 (1997).

\bibitem{501535e337e94fe89cef97bacb43f167}
T.~Krueger {\em et~al.},
\newblock {\em The Lattice Boltzmann Method: Principles and Practice}Graduate
  Texts in Physics (Springer, United Kingdom, 2016).

\bibitem{bird1994molecular}
G.~A. Bird,
\newblock {\em Molecular gas dynamics and the direct simulation of gas flows}
  (Oxford university press, 1994).

\bibitem{Dufty2994}
J.~W.~D. $/*$ and M.~H. Ernst,
\newblock Exact short time dynamics for steeply repulsive potentials, Molecular
  Physics {\bf 102}, 2123 (2004).

\bibitem{PhysRevE.49.1251}
S.~Kambayashi and Y.~Hiwatari,
\newblock Molecular-dynamics study of dynamical properties of dense soft-sphere
  fluids: The role of short-range repulsion of the intermolecular potential,
  Phys. Rev. E {\bf 49}, 1251 (1994).

\bibitem{10.1063/1.3266845}
D.~M. Heyes, S.~M. Clarke, and A.~C. Bra^^c5^^84ka,
\newblock Soft-sphere soft glasses, The Journal of Chemical Physics {\bf 131},
  204506 (2009).

\bibitem{10.1063/1.1730376}
B.~J. Alder and T.~E. Wainwright,
\newblock Studies in molecular dynamics. i. general method, The Journal of
  Chemical Physics {\bf 31}, 459 (1959).

\bibitem{ALLEN1989301}
M.~Allen, D.~Frenkel, and J.~Talbot,
\newblock Molecular dynamics simulation using hard particles, Computer Physics
  Reports {\bf 9}, 301 (1989).

\bibitem{10.1063/1.858656}
H.~A. Hassan and D.~B. Hash,
\newblock A generalized hard-phere model for monte carlo simulation, Physics of
  Fluids A: Fluid Dynamics {\bf 5}, 738 (1993).

\bibitem{1997529}
H.~MATSUMOTO,
\newblock Test of efficiency of variable soft sphere and variable hard sphere
  molecular models in the direct simulation monte carlo method, JSME
  International Journal Series B {\bf 40}, 529 (1997).

\bibitem{Isobe01112016}


\bibitem{Wassgren2006}
C.~Wassgren and J.~S. Curtis,
\newblock The application of computational modeling to pharmaceutical materials
  science, MRS Bulletin {\bf 31}, 900 (2006).

\bibitem{BUIST2016363}
K.~Buist, L.~Seelen, N.~Deen, J.~Padding, and J.~Kuipers,
\newblock On an efficient hybrid soft and hard sphere collision integration
  scheme for dem, Chemical Engineering Science {\bf 153}, 363 (2016).

\bibitem{Zhou2024}
L.~Zhou, M.~A. Elemam, R.~K. Agarwal, and W.~Shi,
\newblock {\em Discrete Element Method (DEM)} (Springer Nature Switzerland,
  Cham, 2024), pp. 83--102.

\bibitem{grad1958principles}
H.~Grad,
\newblock Principles of the kinetic theory of gases,
\newblock in {\em Thermodynamik der Gase/Thermodynamics of Gases}, pp.
  205--294, Springer, 1958.

\bibitem{Boltzmann1964}
L.~Boltzmann, S.~G. Brush, and N.~L. Balazs,
\newblock {Lectures on Gas Theory}, Physics Today {\bf 17}, 68 (1964).

\bibitem{BALMER2011727}
R.~T. Balmer,
\newblock Chapter 18 - introduction to statistical thermodynamics,
\newblock in {\em Modern Engineering Thermodynamics}, edited by R.~T. Balmer,
  pp. 727--762, Academic Press, Boston, 2011.

\bibitem{BIRD1993442}
J.~O. Bird and P.~J. Chivers,
\newblock 58 - the kinetic theory of matter,
\newblock in {\em Newnes Engineering and Physical Science Pocket Book}, edited
  by J.~O. Bird and P.~J. Chivers, pp. 442--448, Newnes, 1993.

\bibitem{Cundall1979}
P.~A. Cundall and O.~D.~L. Strack,
\newblock A discrete numerical model for granular assemblies,
  G^^c3^^a9otechnique {\bf 29}, 47 (1979).

\bibitem{doi:10.1080/19648189.2008.9693050}
S.~Luding,
\newblock Introduction to discrete element methods, European Journal of
  Environmental and Civil Engineering {\bf 12}, 785 (2008).

\bibitem{DANBY2013211}
M.~Danby, J.~Shrimpton, and M.~Palmer,
\newblock On the optimal numerical time integration for dem using hertzian
  force models, Computers \& Chemical Engineering {\bf 58}, 211 (2013).

\bibitem{gray1984theory}
C.~G. Gray, K.~E. Gubbins, and C.~G. Joslin,
\newblock {\em Theory of Molecular Fluids: Volume 2: Applications}volume~2
  (Oxford University Press, 1984).

\bibitem{10.1063/1.437577}
D.~H. Tsai,
\newblock {The virial theorem and stress calculation in molecular dynamics},
  The Journal of Chemical Physics {\bf 70}, 1375 (1979).

\bibitem{Greiner2001}
W.~Greiner,
\newblock {\em The Harmonic Oscillator} (Springer Berlin Heidelberg, Berlin,
  Heidelberg, 2001), pp. 157--183.

\bibitem{PhysRev.159.98}
L.~Verlet,
\newblock Computer "experiments" on classical fluids. i. thermodynamical
  properties of lennard-jones molecules, Phys. Rev. {\bf 159}, 98 (1967).

\bibitem{HART1988117}
R.~Hart, P.~Cundall, and J.~Lemos,
\newblock Formulation of a three-dimensional distinct element model-part ii.
  mechanical calculations for motion and interaction of a system composed of
  many polyhedral blocks, International Journal of Rock Mechanics and Mining
  Sciences \& Geomechanics Abstracts {\bf 25}, 117 (1988).

\bibitem{OTSUBO201767}
M.~Otsubo, C.~O'Sullivan, and T.~Shire,
\newblock Empirical assessment of the critical time increment in explicit
  particulate discrete element method simulations, Computers and Geotechnics
  {\bf 86}, 67 (2017).

\bibitem{PhysRevLett.67.1751}
P.~A. Thompson and G.~S. Grest,
\newblock Granular flow: Friction and the dilatancy transition, Phys. Rev.
  Lett. {\bf 67}, 1751 (1991).

\bibitem{refId01997}
{Christian M. Dury} and {Gerald H. Ristow},
\newblock Radial segregation in a two-dimensional rotating drum, J. Phys. I
  France {\bf 7}, 737 (1997).

\bibitem{LANGSTON1995967}
P.~Langston, U.~Tuzun, and D.~Heyes,
\newblock Discrete element simulation of granular flow in 2d and 3d hoppers:
  Dependence of discharge rate and wall stress on particle interactions,
  Chemical Engineering Science {\bf 50}, 967 (1995).

\bibitem{jccs201500046}
T.~Onishi,
\newblock A molecular orbital analysis on helium dimer and helium-containing
  materials, Journal of the Chinese Chemical Society {\bf 63}, 83 (2016).

\bibitem{Harbola_2011}
V.~Harbola,
\newblock Using uncertainty principle to find the ground-state energy of the
  helium and a helium-like hookean atom, European Journal of Physics {\bf 32},
  1607 (2011).

\bibitem{ZHU201428}
H.~Zhu {\em et~al.},
\newblock Irradiation behaviour of $\alpha_2$ and $\gamma$ phases in he ion
  implanted titanium aluminide alloy, Intermetallics {\bf 50}, 28 (2014).

\end{thebibliography}

\end{document}